# Multiscale analysis of the textural atomization process of a rocket engine assisted coaxial jet

Leonardo Geiger[1,2], Nicolas Fdida[1], Christophe Dumouchel[2], Jean-Bernard Blaisot[2], Luc-Henry Dorey[1], Marie Théron[3]

[1]DMPE, ONERA, Université Paris Saclay, 91123 Palaiseau, France
[2]CORIA, CNRS, Normandie Université, UNIROUEN, INSA ROUEN, 76000 Rouen, France
[3]CNES, Space Transportation Directorate, 75012 Paris, France

**Abstract**
A method for analyzing liquid ligaments of a textural atomization process is presented in this article for the case of a rocket engine type assisted atomization process under combustion. The operating point positions the atomization process in the fiber type regime carrying an intense textural atomization process. Multiscale in nature, the method based on image analysis associates a scale distribution with a family of ligaments, this distribution being sensitive to the number, size and shape of these ligaments. The quality of scale distributions measured by image analysis depends on the spatial resolution and the precision of area measurements of surfaces with curved boundaries but described by square pixels. Part of the work consisted of improving the method for measuring scale distributions by using a sub-pixel image analysis technique and refining the surface area measurement method. Another part directed the multiscale analysis towards the estimation of the diameter distributions of the blobs that characterize the large-scale deformation of the ligaments**.** The analysis describes the atomization process at a level of detail never reached. For instance, assuming that the blobs are drops in formation, the estimated diameter distribution (bimodal in the case examined here) as well as the number of these drops are evaluated as a function of the distance from the injector. This information indicates where the process is most intense and where it stops. Furthermore, these diameter distributions receive a mathematical expression whose parameters report clear evolutions with the distance from the injector. This shows the possibility of elaborating mathematical models appropriate for textural atomization mechanisms.

Corresponding author: Christophe Dumouchel, 0000-0003-0755-4172
christophe.dumouchel@coria.fr, (33)6 32 95 36 23,

Statements and Declarations
The authors have no relevant financial or non-financial interests to disclose.
The authors have no competing interests to declare that are relevant to the content of this article.

Acknowledgments and funding Information
The authors acknowledge CNES and ONERA that funded the research leading to these results.

**Nomenclature**

| | |
|---|---|
| $d$ | scale [L] |
| $d_{pc}$ | modal scale of $e_2(d)_{,d}$ of a cylinder set [L] |
| $D$ | diameter [L] |
| $D_c, D_s$ | mean diameter $D_{q0}$ of a system of cylinders and of spheres respectively [L] |
| $D_{q0}, D_{m0}$ | mean diameter of order $q$ and $m$ order respectively [L] |
| $E_2(d)$ | cumulative scale distribution [L$^2$] |
| $e_2(d)$ | scale distribution [L] |
| $f_{0c}, f_{0s}$ | number-based diameter distribution of a system of cylinders and spheres respectively [1/L] |
| $H$ | length-integrated curvature [-] |
| $L$ | interface length [L] |
| $L_c$ | cylinder length [L] |
| $m(r_I)$ | ): total distance count in bin $r_I$ |
| $N$ | interpolation factor [-] |
| $N_c, N_s$ | number of cylinders and of spheres respectively [-] |
| $n_x, n_y$ | number of pixels in the $x$ and $y$ directions respectively [-] |
| $r$ | exact distance between a pixel and the interface [L] |
| $r_I, r_F$ | integer and fractional part of $r$ respectively [L] |
| $S$ | surface area [L$^2$] |
| $w_{ROI}$ | width of ROI [L] |
| $z$ | distance from the injector exit [L] |
| | |
| $\alpha, q$ | dispersion parameters of the 3pGG distribution [-] |
| $\kappa$ | local curvature [1/L] |
| $\rho_L, \rho_G$ | liquid and gas density respectively [M/L$^3$] |

# 1. Introduction

The atomization of a liquid flow in a gaseous environment is a two-step mechanism referred in the literature as primary and secondary atomization processes. The primary atomization process refers to first events of liquid fragment detachment from the bulk flow and the secondary process concerns the atomization of these fragments.

The primary atomization encompasses two drop production processes involving very different mechanisms that were introduced in a previous publication as textural and structural atomization processes (Dumouchel et al. 2019). Textural atomization refers to the production of liquid fragments and droplets by interface peeling whereas the structural atomization refers to the production of liquid fragments and droplets from the breakup of the liquid bulk. These primary atomization mechanisms produce liquid ligaments and drops of very different size, those emanating from textural atomization being far smaller than the liquid bulk characteristic scale, whereas those produced by structural atomization are more of the order of that very characteristic scale. The work presented in this paper concerns the small-scale process of textural atomization.

Textural atomization processes were reported on large (diameter > 1 mm) round jets either non-turbulent (Hoyt and Taylor, 1977; Wu et al., 1995) or turbulent (Wu et al., 1992, 1993, 1995), on air-assisted non-turbulent or turbulent cylindrical jets (Farago and Chigier, 1992; Marmottant and Villermaux, 2004a), on turbulent liquid sheets (Grout et al., 2007), on cavitating flows (Dumouchel et al., 2019) to quote just a few examples. Experimental studies dedicated to textural atomization processes use imaging approaches to visualize the liquid structures that develop and fragment during the process (Dumouchel (2009) and references therein). Visualizations are performed by shadowgraph and high-speed shadowgraph images, doubled-pulsed holography, LIF tomography, ballistic imaging, X-ray absorption imaging. Being of ever-increasing quality, the images show the liquid structures involved in the atomization processes, the sprays produced and provide fine qualitative descriptions of the mechanism. As for quantitative information, it mainly concerns the sprays (drop size and velocity distribution, more recently information on the shape by image analysis) and little the atomization process itself which is often described by global quantities, the most frequent of which are the breakup length of the jet and the opening angle of the spray at the injector exit. However, we deplore a lack of analysis tools allowing extracting quantitative information related to the shape, number and size of liquid structures involved in the atomization process in order to investigate the origin of drop formation. The present work proposes a method of analysis that tends to fill this gap.

Nevertheless, the studies of the literature report important information on the textural atomization mechanisms. For instance, for the case of non-turbulent large jets, Wu et al. (1995) demonstrated that the boundary layer developed in the injector is required for the textural atomization to occur and that the size of the produced droplets correlated with the thickness of this boundary layer. For large turbulent round jets (Wu et al., 1992, 1993, 1995) the characteristic scales of the drops peeled from the interface are related to the local turbulent scales of the liquid flow. For laminar jets assisted by airflow, the development of textural structures can result from the combination of a Kelvin-Helmholtz instability with a Rayleigh-Taylor instability (Marmottant and Villermaux, 2004a).

An important observation revealed by literature studies is that textural atomization processes always involve the production of ligamentous structures at the liquid-gas interface. Depending on whether these ligaments are produced by the distribution of vorticity due to the boundary layer, by turbulence, by the development of instability, or by any other mechanism, they differ in number, length and shape. The study of the textural atomization processes must therefore rely on the description of atomizing liquid ligaments which leads to the following question: how to describe a population of ligaments representative of a textural atomization process?

The difficulty lies in the fact that the ligaments in question are so deformed that they do not show axial symmetry. The work of Villermaux et al. (2004) proposes to represent a deformed ligament by a series of circular blobs of various diameters and whose diameter distribution provides a description of the ligament size, length and shape. A ligament fragmentation model incorporating this diameter distribution and representing the breakup phase through agglomeration of the blobs was proposed, leading to the mathematical writing of the resulting drop diameter distribution. This mathematical solution was applied with success in several situations (Marmottant and Villermaux, 2004a, 2004b; Villermaux et al., 2004) as well as for ligaments of viscoelastic liquids (Keshavarz et al. 2016). However, the blob diameter distribution describing a ligament is conceptual and not associated with any method for determining it.

The development of textural ligaments modifies the tortuosity of the liquid-gas interface. The fractal analysis is a mean of quantifying this tortuosity. The first to proceed a fractal analysis on atomizing liquid flow are Shavit and Chigier (1995) for the case of a coaxial cylindrical jet. They found a fractal dimension of the interface that increases and decreases as a function of the distance from the nozzle. The location at which the fractal dimension was found maximum is where the breakup became effective. In a previous work, we performed a fractal analysis of an atomizing turbulent liquid sheet (Grout et al., 2007) and found two main fractal dimensions, the first one characterizing the textural atomization process close to the injector and the second one characterizing the structural atomization process further downstream. It was observed the textural fractal dimension correlated with the Reynold number of the issuing liquid flow revealing the connection between the turbulence of the textural atomization process in agreement with the conclusions of the studies of Faeth's group for round jets.

The morphological description provided by the fractal analysis served as a starting point for the development of a multiscale analysis aimed at producing a quantitative description of tortuous interfaces and more particularly those of atomizing liquid flows (Dumouchel, 2017). In a previous investigation (Dumouchel et al., 2019), this multiscale tool was applied to analyze the ligaments of the textural atomization process observed on a cavitating flow. The ligaments were described by a scale distribution measured by image analysis. This distribution contains all information related to the ligaments including their size, number and shape. In addition, that work demonstrated the possibility of associating a mathematical function with this scale distribution, using the fact that any set of ligaments, whatever their number, size and shape, admits an equivalent set of cylindrical ligaments, i.e., a set of cylindrical ligaments that has the same scale distribution. In addition to a description of tortuous interfaces, multiscale analysis opens a way to represent them mathematically.

The present work intends to apply the multiscale analysis to investigate the textural atomization process observed on a coaxial jet involving a central liquid flow and a peripheral gas flow. The operating point of the considered experimental situation reproduces a rocket engine type injection with a fiber type atomization process according to Lasheras and Hopfinger (2000) classification and known to show an intense textural atomization process. The experimental injection setup takes place under combustion, i.e., under severe conditions due to flame luminosity, optical index gradients or particle deposition on windows. Under these conditions, the measurement of drops around the jet by conventional optical techniques (Laser diffraction technique, Particle Phase Doppler Analyzer) cannot be used. The study of the ligamentous structures leading to the production of these drops is an alternative to approach some quantitative elements concerning them.

As far as this last point is concerned, one of the objectives of this work is to propose an analysis of the scale distributions of the textural liquid ligaments in order to determine the size distribution of the blob population that structures the ligament deformation. These blobs, similar to those introduced by Villermaux et al. (2004), are a first indication of drops in production and their size distribution is believed to be related to the diameter distribution of the drops these ligaments will produce. This analysis makes use of recent theoretical advances concerning the multiscale description tool (Dumouchel et al., 2022; Dumouchel et al., 2023).

The study is based on the analysis of high spatial resolution images. Furthermore, thanks to the use of the multiscale approach, quantitative elements such as the size of the ligaments or their number are expected. The multiscale description requires images with a very high spatial resolution. This is all the more important for textural atomization processes that are small-scale mechanisms. Moreover, the representation of curved boundaries with square pixels is a source of errors in the measurement of the scale distribution. Part of the work presented in this article is devoted to developing improved image processing and measurement technique to evaluate and significantly reduce the negative impact of these points on the measurement and analysis of the scale distributions.

The following section gives information on the experimental set-up and the optical diagnostic. Then a section dedicated to the multiscale analysis and to the improved measurement protocol is presented. The last section presents the results of the analysis and the paper ends with a conclusion.

## 2. Experimental setup and optical diagnostic

The ONERA's MASCOTTE test-bench is used for the present experiments. This bench is a subscale cryogenic rocket combustor capable of reproducing operating conditions similar to those encountered inside liquid rocket engine combustion chambers, i.e, high pressures and high mixture ratios. It is equipped with multiple sensors, as well as large optical access windows allowing the use of non-intrusive diagnostics and the investigation of several mechanisms including injection, atomization and combustion of the propellants (Habiballah et al. 2006, Fdida et al. 2016, Boulal et al. 2022), onset of combustion instabilities, as well as heat transfer (Grenard et al. 2019), and pollutant and soot formation.

In the present work, the injection system is a single shear-coaxial injector without recess between the liquid and the gas injector-tube exit plane. It injects a cylindrical liquid jet of oxidizer (liquid oxygen) from the central circular tube surrounded by a high-speed gaseous fuel (methane) flow injected from the outer annular tube. The oxidizer-fuel mixture is slightly fuel-rich, with a near-stoichiometric mixture ratio as commonly encountered in liquid rocket engine main combustion chamber nominal operation (Preuss et al. 2008). The propellants are injected in a subcritical state, i.e., either their pressure or temperature are below their critical point values (Lux and Haidn, 2009). A single injection operating condition is used in this work. It is characterized by a chamber pressure equal to 7 bar, mixture ratio equal to 3.5, a liquid Reynolds number equal to 77 $10^3$, a relative gaseous Weber number equal to 46 $10^3$ (both based on the cylindrical jet diameter) and a momentum flux ratio equal to 14.5. According to Lasheras and Hopfinger (2000) classification, these figures place the atomization process in the fiber-type breakup regime.

Injection takes place under combustion conditions. The methane-oxygen mixture is ignited with a hydrogen-oxygen torch-igniter. Under such subcritical combustion, the dense mist surrounding and masking the liquid jet is rapidly consumed and the liquid jet becomes visible again (Fdida et al. 2016). This is illustrated in Fig. 1 where a large view of the jet in the vicinity of injection face plate is shown. This reveals shimmering areas around the liquid jet that are the manifestation of optical index gradients caused by the strong temperature gradients resulting from combustion.

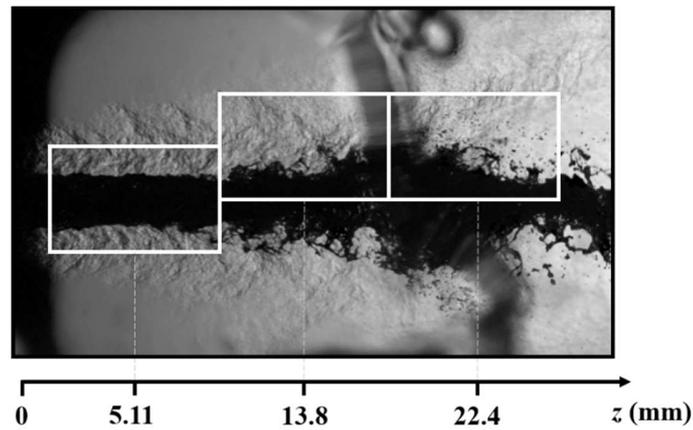

**Figure 1:** Positions of the three-image series
($z = 0$ corresponds to the position of the injector exit section)

The combustion chamber is equipped with two opposite rectangular quartz optical windows of 25 mm x 60 mm to record images by adopting a backlighting optical arrangement. The imaging system is composed of a pulsed laser source illuminating the flow field through the combustion chamber. This source (Cavilux Smart 400 W) emits a red incoherent light pulse at a wavelength of 640 ± 10 nm. The pulse duration is equal to 1-1.5 µs which is short enough to ensure the freeze of the jet. A diffuser is placed at the fiber exit, and the focal point of a Fourier lens is placed at the diffuser position to create a collimated and homogeneous illumination background. In the area close to the injector, optical index gradients can be seen on instantaneous images. These gradients are due to the presence of the flame, but also to the turbulence and gas composition (Fdida et al. 2019). They are suppressed by the segmentation step mentioned in section 3.2. The camera used is a JAI GOX-2402 with 1920 x 1200 pixels and a pixel size = 3.45 µm. The spatial resolution and the position of the imaging system is obtained thanks to a calibrated target with known dimensions positioned at a given distance from the injector exit. Reference images are recorded prior to the test run to obtain first the pixel to millimeter conversion and second the image plane coordinates in the reference frame of the test chamber.

With the aim of visualizing small-scale interfacial liquid structures and describing them using a multiscale approach, the highest possible spatial resolution is required. The camera is therefore equipped with a long-distance microscope Infinity K2 DistaMax with a CF1/B lens. This setup results in a field of view of 8.64 mm x 5.40 mm (white rectangles in Fig. 1) with a spatial resolution of 4.5 µm/pixel. Finally, a narrow band filter is used to block the light from the flame emission while allowing the light from the laser source to reach the sensor.

In order to investigate the spatial evolution of the textural atomization process, image series centered on three distances from the injector were recorded. The positions of the image series are shown in Fig. 1. Since the liquid jet expanded radially while atomizing, the image series at the two farthest positions were off-centered, reducing viewing to a single interface, unlike the series of images at $z = 5.11$ mm (see Fig. 1). At each position 100 images were recorded during the stationary phase of the reactive flow.

### 3. Multiscale analysis – Description and Measurement
#### 3.1 Multiscale description

The description of the textural deformation of the liquid gas interface in the injector near field region uses the multiscale analysis of the visualized liquid system. Introduced and applied in several previous works (Thiesset et al. 2019, Dumouchel et al. 2017, 2019, 2022, 2023) this analysis considers the liquid system as well as all systems parallel to it in the liquid phase. The parallel systems are those obtained by erosion operations. Such operation is a classical image analyzing tool that consists in removing the liquid system region covered by a disk of diameter $d$ that was dragged along the interface keeping the disk center on the interface line. The remaining system is the system eroded at scale $d$. As an example, a disk of diameter $D$ (Fig. 2-a) eroded at scale $d$ (Fig. 2-b) is a disk of diameter $(D – d)$ (Fig. 2-c). Since both disks have the same center, they are said parallel with each other. The scale $d = D$ is the maximum scale of the system, i.e., the smallest scale that leads to the total erosion of the initial system. In the present application, the erosion operation is performed at all possible scales, i.e., until the initial liquid system has been totally eroded.

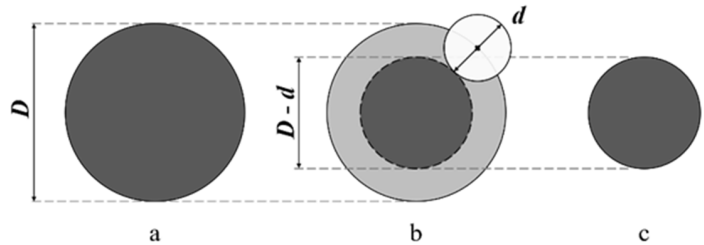

**Figure 2:** a – Disc of diameter $D$, b – Eroded operation at scale $d$,
c – The eroded system at scale $d$ is a disc of diameter $D – d$.

Since the embedded dimension of the analysis is 2, the liquid system and its parallel systems can be described by three integral quantities, i.e., their surface area $S(d)$, their interface length $L(d)$ and their length-integrated-curvature $H(d)$:

$$H(d) = \int_{L(d)} \kappa(d) dl \tag{1}$$

where $\kappa(d)$ is the local curvature of the eroded system at scale $d$. $S(0)$, $L(0)$ and $H(0)$ characterize the surface area delimited by the liquid gas interface and the length and length-integrated-curvature of this interface, respectively. Previous works (Dumouchel et al. 2022, 2023) have established the expressions linking the functions $S(d)$, $L(d)$ and $H(d)$. Through the $d$-scale, they reveal a dependence on the shape of the systems.

In the present work, the liquid system and their parallel systems are described by the cumulative scale distribution $E_2(d)$. The function $S(d)$ allows to draw this distribution that is defined as $E_2(d) = S(0) - S(d)$. $E_2(d)$ increases from 0 to $S(0)$ when $d$ increases from 0 to the maximum scale. The derivative of $E_2(d)$ with respect to the scale $d$ is the scale distribution $e_2(d)$: $e_2(d) = E_2(d)_{,d}$ where $\bullet_{,d}$ denotes the derivative with respect to $d$. As demonstrated elsewhere (Dumouchel et al. 2022, 2023) $e_2(d) = L(d)/2$.

Furthermore, it was also shown that the absence of tips on the liquid gas interface due to the action of surface tension forces leads to:

$$-e_2(0)_{,d} = \frac{H(0)}{4} \tag{2}$$

The property expressed by Eq. (2) for $d = 0$ remains valid as long as the erosion scale $d$ is small enough to produce a tip-free eroded-system. As illustrated in Fig. 3 for the case of an ellipse (Fig. 3-a), eroded systems on small scales retain a smooth boundary (Fig. 3-b), while those on larger scales may exhibit tips (Fig. 3-c). Equation (2) can be generalized for small scales only.

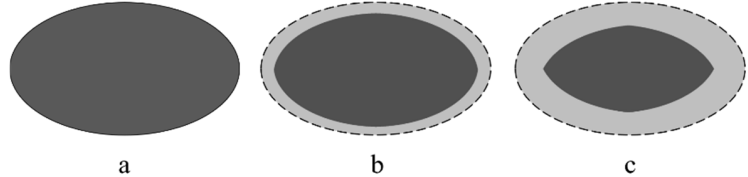

**Figure 3:** a – Ellipse, b – Ellipse eroded at a small scale: the eroded system retains a smooth boundary, c – Ellipse eroded at a large scale: the eroded system shows two tips.

The analysis carried out in this work is based on the study of the distribution $-e_2(d)_{,d}$. To help with the analysis, a theoretical set of cylinders whose diameters $D$ are distributed according to a given distribution is considered. This synthetic set is of interest because, as demonstrated in a previous study (Dumouchel et al. 2019), any set of deformed ligament structures can be associated with an equivalent set of cylinders, i.e. a set with the same scale distribution. Thus, the system studied can be characterized by a diameter distribution, that corresponds to the equivalent system of cylinders.

We consider a set of $N_c$ cylinders with the same length $L_c$ but whose diameters $D$ are distributed according to a number-based distribution $f_{0c}(D)$. The 2D segmented image of a cylinder of diameter $D$ being a black band of width $D$ with no end on a white background. The interface length carried by cylinders with diameters between $D$ and $D + \Delta D$ is $\Delta L = 2N_cL_cf_{0c}(D) \Delta D$ and, since $e_2(d) = L(d)/2$:

$$-e_2(d)_{,d} = N_cL_cf_{0c}(D)_{D=d} \qquad (3)$$

Equation (3) reveals two important points for systems composed of cylinders or/and of ligamentous structures associated with an equivalent system of cylinders:
1. The distributions $f_{0c}(D)$ and $-e_2(d)_{,d}$ show as many modes as each other. (If the mode is single, as is the case in this work, the scale of this mode is noted $d_{pc}$)
2. Since by definition $f_{0c}(0) = 0$ then $-e_2(0)_{,d} = 0$.

In a real atomizing liquid system, the textural structures are ligamentous but not purely cylindrical: they can be seen as cylinders showing successive contracted and dilated sections. The distribution $-e_2(d)_{,d}$, and thus the equivalent-system diameter distribution $f_{0c}(D)$, reflect the distribution of these section scales. Small scales mainly characterize pinching whereas large scales mainly characterize swelling.

As far as the present analysis is concerned, two mathematical elements are introduced.

1 – As done by Dumouchel et al. (2019), the cylinder diameter distribution in Eq. (3) is represented by a three parameter Generalized-Gamma (3pGG) function which is a mono-modal distribution (Dumouchel, 2006). Thus, Eq. (3) becomes:

$$-e_2(d)_{,d} = N_cL_c \frac{q}{\Gamma\left(\frac{\alpha}{q}\right)}\left(\frac{\alpha}{q}\right)^{\frac{\alpha}{q}}\frac{d^{\alpha-1}}{D_c^\alpha}\exp\left(-\frac{\alpha}{q}\left(\frac{d}{D_c}\right)^q\right) \qquad (4)$$

where $q$, $\alpha$ are positive real numbers, $D_c$ is a mean diameter ($D_c$ is actually equal to the mean diameter $D_{q0}$ of the cylinder diameter distribution), and $\Gamma$ is the Gamma function (Abramowitz and Stegun, 1964). The modal scale of this distribution is:

$$d_{pc} = \left(\frac{\alpha-1}{\alpha}\right)^{\frac{1}{q}} D_c \qquad (5)$$

2 - When the ligament structures are fully atomized as an ensemble of drops, the liquid system shows no pinching anymore and it was shown that $d_{pc} = 0$ (Thiesset et al., 2019). According to Eq. (5), this value of $d_{pc}$ is obtained with $\alpha = 1$. In this case, Eq. (4) becomes:

$$-e_2(d)_{,d} = \frac{N_cL_c}{D_c}\frac{q^{(1-1/q)}}{\Gamma\left(\frac{1}{q}\right)}\exp\left(-\frac{1}{q}\left(\frac{d}{D_c}\right)^q\right) = -e_2(0)_{,d}\exp\left(-\frac{1}{q}\left(\frac{d}{D_c}\right)^q\right) \qquad (6)$$

Therefore, a mathematical model for the size distribution of drops (which are assumed spherical) can be proposed. Indeed, the diameter distribution $f_{0s}(D)$ of a set of spherical drops is proportional to the second derivative of the scale distribution $e_2(d)$ (Thiesset et al. 2019). According to Eq. (6) this leads to:

$$-e_2(d)_{,d,d} = \frac{N_c L_c}{D_c} \frac{q^{(1-1/q)}}{\Gamma\left(\frac{1}{q}\right)} \left(f_{0s}(D)\right)_{D=d} \tag{7}$$

where the normalized distribution $f_{0s}(D)$ is given by:

$$f_{0s}(D) = \frac{D^{q-1}}{D_c^q} \exp\left(-\frac{1}{q}\left(\frac{D}{D_c}\right)^q\right) \tag{8}$$

The mean diameter series $D_{m0}$ and the modal diameter $D_{ps}$ of this distribution are:

$$\begin{cases} D_{m0} = q^{\frac{1}{q}} \left[\Gamma\left(\frac{m}{q}+1\right)\right]^{\frac{1}{m}} D_c \\ D_{ps} = (q-1)^{\frac{1}{q}} D_c \end{cases} \tag{9}$$

Thanks to this equation, the number-based diameter distribution $f_{0s}(D)$ can be rewritten as:

$$f_{0s}(D) = \frac{qD^{q-1}}{D_s^q} \exp\left(-\left(\frac{D}{D_s}\right)^q\right) \tag{10}$$

where $D_s$ is the mean diameter $D_{q0}$ of the distribution $f_{0s}(D)$. According to Eq. (9), $D_s = q^{1/q} D_c$. Eq. (10) says that $f_{0s}(D)$ is also a 3pGG distribution with $\alpha = q$. Furthermore, Eq. (9) reveals that $D_{10} \approx D_{ps}$ when $q$ is greater than 2.5, saying that the distribution can be considered as symmetric in this case. Finally, considering the expression of the second derivative of the scale distribution $e_2(d)$ for a set of spheres, i.e. (Thiesset et al. 2019):

$$-e_2(d)_{,d,d} = \frac{\pi}{2} N_s \left(f_{0s}(D)\right)_{D=d} \tag{11}$$

where $N_s$ is the number of spheres, Eq. (7) indicates that:

$$\frac{\pi}{2} N_s = \frac{N_c L_c}{D_c} \frac{q^{(1-1/q)}}{\Gamma\left(\frac{1}{q}\right)} \tag{12}$$

3.2 Image processing for scale distribution measurements

The images undergo a certain amount of processing to produce segmented images in which the liquid appears in black on a white background. Prior to segmentation, a normalization step to reduce local lighting inhomogeneities is performed by dividing each image by a background image (image of the background with no jet). The subsequent segmentation step uses the ImageJ local Phansalkar thresholding algorithm (Phansalkar et al., 2011) with a square local analysing window of 135 µm side length. The liquid structures detached from the liquid core are eliminated and the holes in the liquid phase are filled with black pixels. An illustration of these image processing steps is available in Dumouchel et al. (2019).

The surface areas $S(d)$ of the liquid system and its parallel systems are measured on the segmented images applying an Euclidian Distance Map (EDM) operation. For each liquid pixel, this operation determines the shortest distance between said pixel and the interface: a brightness level equal to this distance is attributed to each liquid pixel. The surface areas $S(d)$ correspond to the number of pixels whose brightness level is equal to or larger than $d/2$. In this exercise, the initial scale value is $d = 0$ and the scale increment $\Delta d = 2$ pixels, since $d$ represents the diameter of the erosion circle. Considering that the spatial resolution is equal to 4.5 µm/pixel (see Section 2), $\Delta d = 9$ µm. The surface areas $S(d)$ allows determining the cumulative scale $E_2(d)$ and successive derivatives $e_2(d)$ and $e_2(d)_{,d}$ introduced above. (The derivative calculation step is always preceded by a smoothing of the function to be derived. This smoothing consists of averaging over five consecutive points). The EDM algorithm applied is the ImageJ *Distance map function* yielding a 32-bit grayscale EDM-transformed image.

As an illustration, a measurement is applied on the synthetic image shown in Fig. 4 that represents a 70 pixels diameter cylinder disturbed by a sinusoidal perturbation of wavelength 800 pixels and amplitude 31 pixels. A theoretical expression of the distribution $-e_2(d)_{,d}$ for such a system is available elsewhere (Dumouchel et al., 2023).

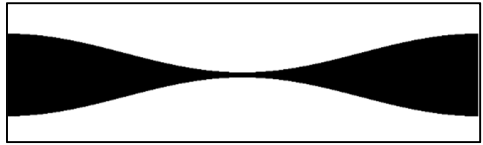

**Figure 4:** Synthetic image of a sinusoidally perturbed cylinder
(unperturbed diameter: 70 pixels; image width and perturbation wavelength: 800 pixels; perturbation amplitude: 31 pixels)

The theoretical and measured distributions $-e_2(d)_{,d}$ of the system shown in Fig. 4 are compared with each other in Fig. 5-a. The theoretical result shows two discontinuities, one at $d_1 = 8$ pixels corresponding to the diameter of the cylinder contraction and one at $d_2 = 132$ pixels corresponding to the diameter of the cylinder swelling (see Dumouchel et al., 2023). The measurement in Fig. 5-a shows average agreement with this theoretical solution. The scales of discontinuities are poorly understood, especially the small one, and the measurement shows oscillations between these two scales. Finally, we note that the measurement does not tend towards zero when $d$ decreases towards zero, as theoretically expected (see Eq. (3)). The objective of this study being to measure small-scale liquid features, these differences must be addressed.

One of the sources of error comes from the fact that ImageJ *Distance map function* assigns to each liquid pixel a distance to the interface that always corresponds to an integer number of pixels. As a consequence, the surface areas $S(d)$ are also integer numbers of pixels which induces a bias in the measurement. The following two step process intends to correct this bias.

First, the exact distances between the liquid pixels and the interface are calculated by using the Python *SCIPI.NDimage.distance_transform_EDT* routine (which actually is the same as the ImageJ *Exact.Distance.Transform* routine). The exact distance $r$ is defined as:

$$r = \sqrt{n_x^2 + n_y^2} \qquad (13)$$

where $n_x$ and $n_y$ correspond to numbers of pixels as shown in Fig. 6.

According to Eq. (13), the distances $r$ constitute a series of discrete values. They can be rewritten as:

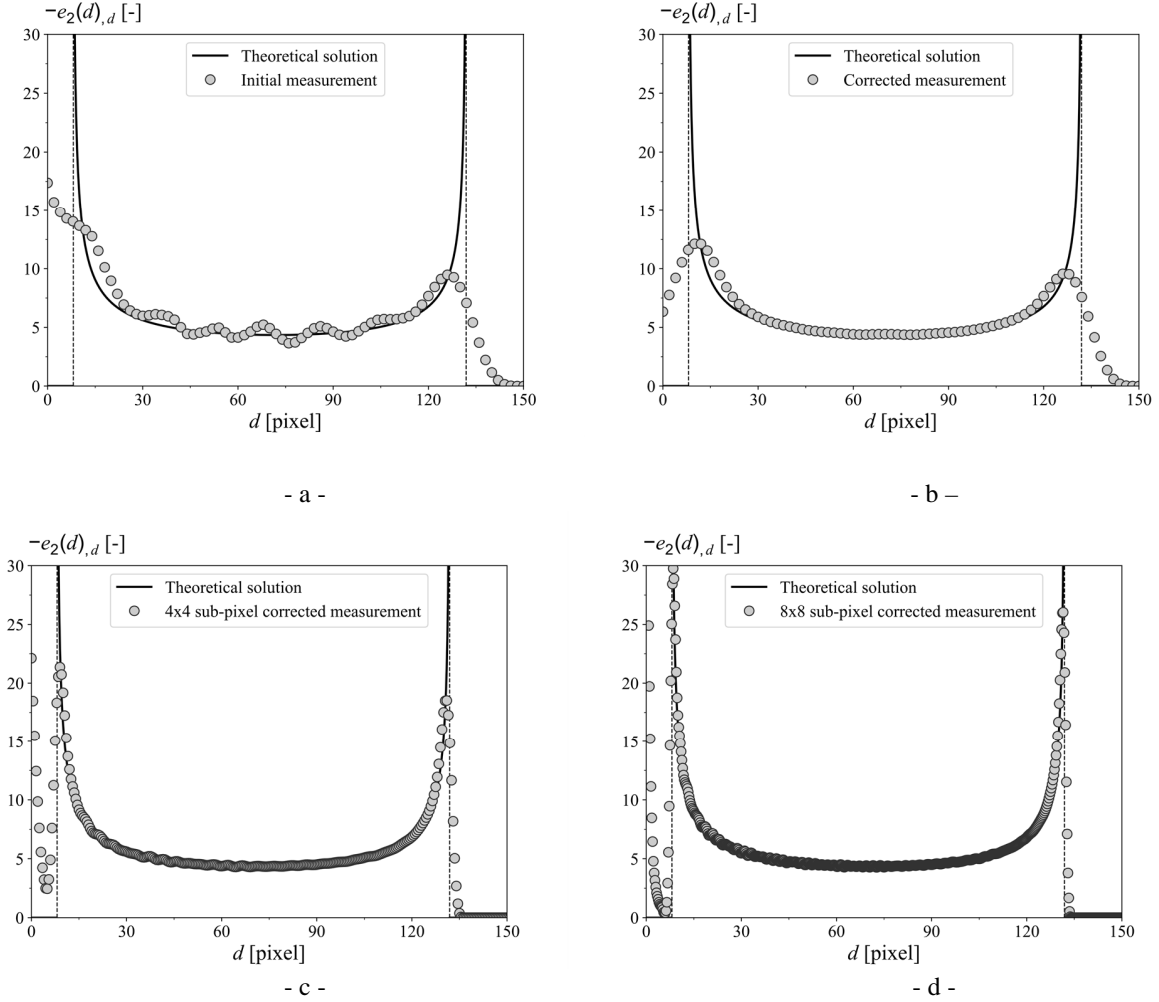

**Figure 5:** Comparison between the theoretical and the measured distribution $-e_2(d)_{,d}$
a – Initial measurement, b – Corrected measurement
c – Corrected measurement with 4x4 subpixel interpolation
d - Corrected measurement with 8x8 subpixel interpolation

$$r = r_I + r_F \qquad (14)$$

where $r_I$ and $r_F$ designate the integer and fractional parts of $r$, respectively.

Second, the exact distance series $r$ (Eq. (14)) is distributed in a series of bins, each of which corresponds to an integer distance $r_I$. The repartition of the exact distance series $r$ in the integer distance series $r_I$ is performed according to the following proportion:

$$\begin{cases} (1 - r_F) \text{ in bin } r_I \\ r_F \text{ in bin } (r_I+1) \end{cases} \qquad (15)$$

Each liquid pixel whose distance from the interface has an integer part equal to $r_I$ feeds both the $r_I$ and $r_I + 1$ distance bins according to Eq. (15). Two examples of this redistribution are presented in Fig. 6. Thus, the surface area of the parallel system resulting from a $d$-scale erosion, $S(d)$, can be expressed as:

$$S(d) = S(0) - \sum_{r_I=0}^{d/2} m(r_I) \qquad (16)$$

where $m(r_I)$ corresponds to the total distance count in bin $r_I$, i.e. when the distance contribution of every pixel in the system is considered.

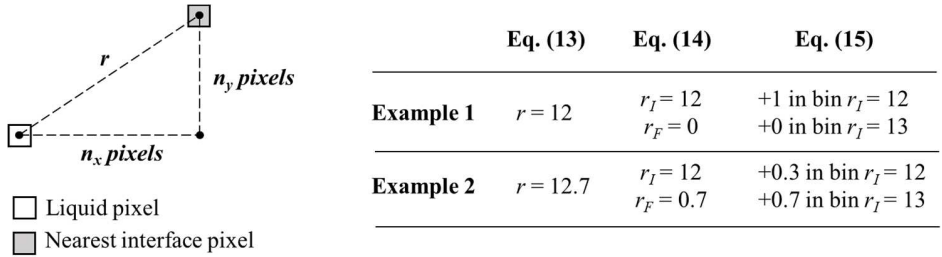

**Figure 6:** Definition of $n_x$ and $n_y$ introduced in Eq. (13). Examples of applications of Eqs. (13) to (15).

The surface areas expressed by Eq. (16) are no longer necessarily equal to an integer number of pixels and this has a positive repercussion on the determination of $-e_2(d)_{,d}$ as shown in Fig. 5-b where we note that the measurement oscillations obtained before correction have completely disappeared. However, even though the positions of the maxima of the measured function correspond satisfactorily to the theoretical peaks, the measured peaks remain much wider than the theoretical ones and the expected theoretical behavior (i.e. $-e_2(d)_{,d} = 0$) for scales smaller than $d_1$ is far from being retrieved by the measurement. Caused by an insufficient image spatial resolution, this disagreement is mitigated by adopting a sub-pixel image analysis.

The sub-pixel treatment consists in applying a bilinear interpolation on the pixel gray-level values to produce sub-pixel segmented images (Fadnavis, 2014). A $N \times N$ interpolation corresponds to a pixel size divided by $N$ and to a spatial resolution multiplied by $N$. Figures 5-c and 5-d show the measured distribution $-e_2(d)_{,d}$ after a 4 x 4 and 8 x 8 sub-pixel interpolation, respectively. These two examples show that sub-pixel analysis considerably improves measurement as far as the determination of the peaks of the distribution are concerned. There is still a peak located at scale $d = 0$ testifying to the inevitable persistence of effects linked to image pixelization. In the examples shown in Fig. 5, the dissociation between the peak of the distribution in the small scale and the peak at $d = 0$ allows to determine the smallest structure scale correctly detectable by the procedure. This scale is equal to 5 pixels here.

This section ends with examples of measured distribution $-e_2(d)_{,d}$ on experimental images. First, the number of images required to have statistically converged distributions is determined by plotting the average distribution as a function of the number of images (Fig. 7-a). We note that the 100 images available at each position are sufficient to have relevant averaged distributions.

Second, the advantage brought by the subpixel treatment is illustrated (Fig. 7-b). The high number of images limited the level of the sub-pixel treatment so that it remained feasible. Thus, the 4 x 4 sub-pixel treatment was applied only here as well as in all the results presented in this article. Without subpixel treatment, the spatial resolution is 4.5 µm/pixel, and, as mentioned earlier, $\Delta d = 9$ µm. With a 4x4 sub-pixel treatment, the spatial resolution reduces to 1.125 µm/pixel and $\Delta d = 2.25$ µm. In Fig. 7-b, the results are displayed in the scale range [0 µm; 250 µm]. In agreement with Eq. (3), the distribution $-e_2(d)_{,d}$ is expected to display a peak. Both distributions in Fig. 7-b show this for the same scale. As explained above, the supplementary peak at $d = 0$ obtained for the distribution issued from the subpixel analysis results from the persistent pixelization bias. According to Eq. (3), it is also expected that $-e_2(0)_{,d} = 0$. Of the two distributions shown in Fig. 7-b, the peak of the one obtained with subpixel analysis is close to satisfying this property. Therefore, the peak reported by the subpixel analysis is believed to be more accurate and, in particular, the part of the distribution to the right of the peak. It is precisely on this part of the distribution that the analysis presented below focuses. Figure 7-b also shows that the smallest structure-scale the procedure can detect is of the order of 16 µm.

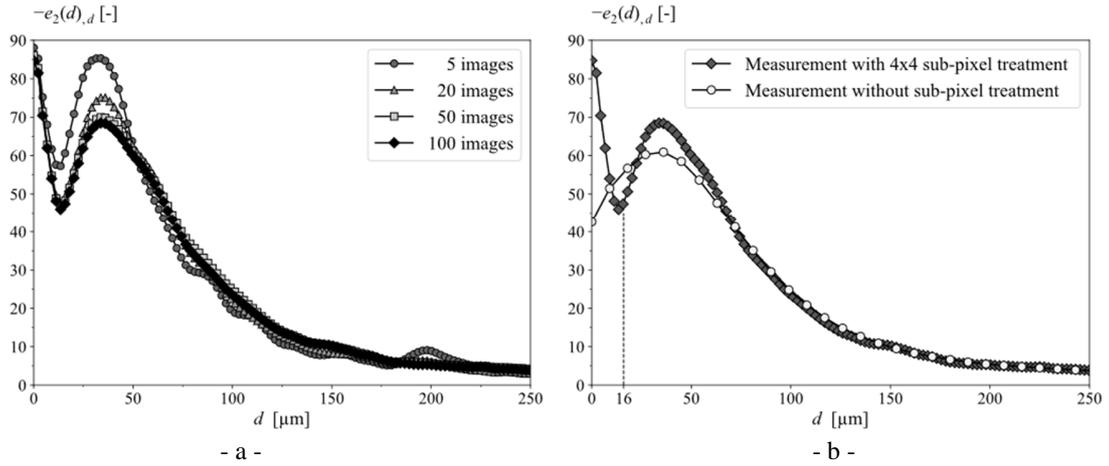

**Figure 7:** Measurement of $-e_2(d)_{,d}$ at $z = 13.8$ mm (see Fig. 1)
a – Influence of the number of images (measurement with subpixel treatment),
b – Illustration of the subpixel treatment.

## 4. Results and Analysis

A series of segmented images of the oxygen jet upper interface in the near injector nozzle region are shown in Fig. 8 as a function of the position $z$. These images, in which the liquid appears in black, are not correlated in time.

The liquid-gas interface shows large-scale and small-scale deformations. The large-scale deformation can be likened to wave with a wavelength of the order of a few millimeters. They are probably the result of the development of a Kelvin-Helmholtz instability commonly encountered in this type of flow.

Small-scale deformations are what make up the textural atomization process. Ligamentous in shape, these liquid protrusions produce droplets at the periphery of the liquid jet (as mentioned above, these drops were actually erased and are not visible in Fig. 8). This figure shows that the size, length, number and shape of these ligaments evolve significantly with the distance $z$. As $z$ increases, the ligaments become more numerous, longer, larger and increasingly deformed. This observation is similar to that reported by Wu et al. (1992) in their study devoted to the primary breakup in gas/liquid mixing layer for turbulent liquids. In line with the conclusions of their work, we may infer that the textural ligaments are initiated by the turbulent fluctuation in the liquid. At each position, the development of the ligaments is associated with conditions in which the momentum of turbulent fluctuations in the liquid is sufficient to overcome surface tension forces. As we move away from the injector, the characteristic size of the turbulent fluctuations that meet this condition increases, leading to the production of thicker ligaments. Therefore, these ligaments will likely produce larger droplets.

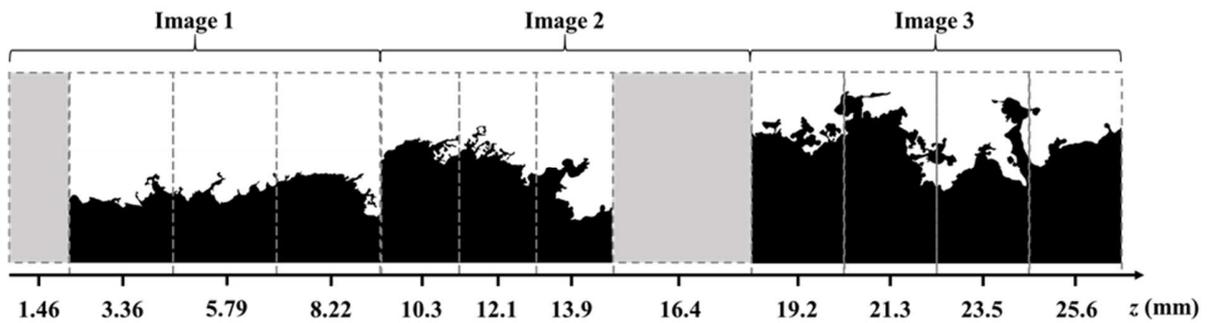

**Figure 8:** Snapshot of the liquid jet upper interface as a function of the distance from the injector exit (Images after treatment, Liquid appears in black, Position of the ROI).

We see also in Fig. 8 that the textural ligaments are mainly oriented in the streamwise direction, particularly near the injector (images 1 and 2). This denotes aerodynamic effects. According to Wu and Faeth (1993), the primary break-up regime (i.e. the textural process) is affected by aerodynamics when the density ratio $\rho_L/\rho_G$ is less than 500. It is equal to 224 in the present study. Furthermore, the present liquid flow is surrounded by a high velocity gas flow explaining the orientation of the textural ligaments. Near the injector, the aerodynamic effect is associated with effects of pressure drops caused by the acceleration of the surrounding gas over protrusions from the liquid surface (Wu and Faeth, 1993). According to Wu and Faeth (1993) the aerodynamic force effects further downstream is to trigger a secondary breakup of the drops produced by the textural ligaments.

The present work focuses on the description of textural ligaments with the aim to estimate the size of the drops these ligaments may produce. The estimation of these drops is based on the description of the shape of these ligaments.

An atomizing ligament reveals areas of swelling which are most likely the signature of drops in production. These areas are schematized by circular blobs in Fig. 9. (The use of circular blobs to describe atomizing ligaments is reminiscent of the work of Villermaux et al. (2004).) The analysis presented in this section aims at estimating the size distribution of these blobs and to consider that this size distribution is indicative of the diameter distribution of the large drops the ligaments will produce.

The determination of the textural blob size distributions is achieved by analyzing the interface scale distributions. These distributions are measured on the Regions of Interest (ROI) that are shown in Fig. 8. This spatial division gives access to the spatial evolution of the blob size distribution. The reduction of the analyzing zone imposed by the ROI is possible without affecting the accuracy of the measurement because, as shown in Fig. 7-a, 100 images are far enough to have converged averaged distribution. The scale distributions reported hereafter are averaged of 100 measured distributions. In Fig. 8, the left zone of Image 1 (gray zone centered on $z = 1.46$ mm) could not be considered in the analysis since, as shown in Fig. 1, the liquid flow is not visible there. Furthermore, because of deposition on the windows, the right zone of Image 2 (gray zone centered on $z = 16.4$ mm in Fig. 8) could not be analyzed either. Therefore, the ROI in images 1, 2 and 3 are different in number and in width. Images 1 and Images 2 are divided in three ROI of width $w_{ROI} = $ 2.43 mm and 1.79 mm, respectively. Images 3 are divided in four ROI of width $w_{ROI} = 2.16$ mm.

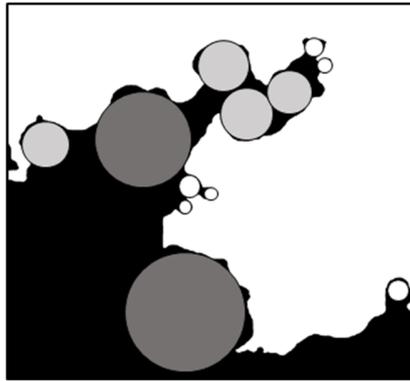

**Figure 9:** Identification of circular blobs on a textural ligament

Figure 10 plots the scale distribution $e_2(d)$ for all positions. In order to make the results comparable with each other, these distributions were divided by the ROI width $w_{ROI}$. The distributions measured on the image 1 ROIs have been divided by 2, as they were measured on two interfaces. Finally, the distributions are plotted in the scale range [16 µm; 250 µm] only since liquid structures of smaller scale can't be spatially resolved (see previous section).

As explained above, the scale distribution $e_2(d)$ represents the half of the circumference of the parallel system series. For $d = 0$, this circumference is equal to the interface length.

First, Fig. 10 shows that the scale distributions decrease with the scale $d$. Observed at each $z$ position, this decrease is representative of the size and deformation of the ligamentary structures: the finer the structures and their deformation, the smaller the scales at which the distribution decays. Second, we note that the distributions reach a constant value as the scale increases. The scale at which this happens corresponds to the largest scale of the textural liquid ligaments.

When the position $z$ increases, we see that the distribution decays spread on larger scales revealing an increase of the size of the textural ligaments. This is consistent with the observations made in Fig. 8. Furthermore, the distribution $e_2(d)$ tends to increase when the position $z$ ranges from 3.36 mm to 21.3 mm and then decreases for the two farthest positions. In a way, $e_2(d)$ contains information on the volume of the textural liquid structures. Thus, an increase in $e_2(d)$ at a given scale can be associated with an increase in the volume carried by these structures. From this point of view, we deduce that the volume of liquid involved in the textural atomization process initially increases and then decreases when the distance from the injector increases.

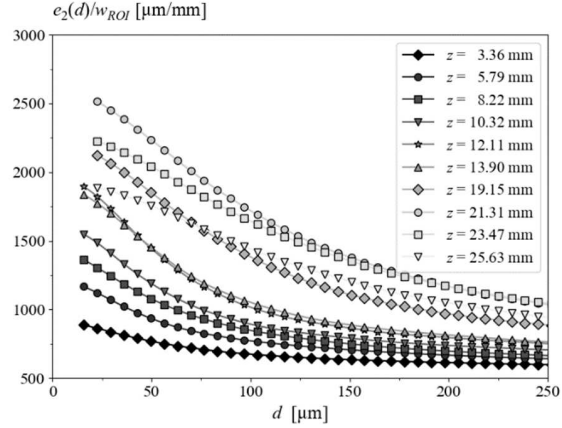

**Figure 10:** Distribution $e_2(d)/w_{ROI}$ for all $z$ positions
(For clarity, the curves show every third point)

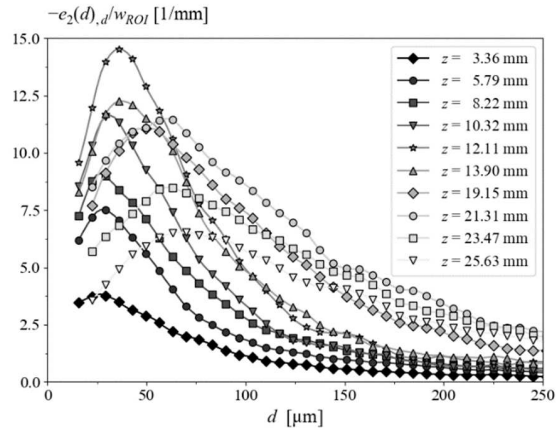

**Figure 11:** Distribution $-e_2(d)_{,d}/w_{ROI}$ for all $z$ positions
(For clarity, the curves show every third point)

The derivatives $-e_2(d)_{,d}$ of the scale distribution $e_2(d)$ are plotted in Fig. 11. As expected, these distributions show bell shapes that are proportional to the number-based diameter distribution $f_{0c}(D)$ of the equivalent set of cylinders (see Eq. (3)). By imposing $L_c = 1$ µm (which is done throughout the analysis), the proportionality coefficient is equal to the number of elements of that length that contribute to this distribution.

As mentioned above, the right-hand side of the distribution (for scales greater than $d_{pc}$) mainly contains the swelling scales. It is therefore in this part of the distribution that the information concerning the blob size distribution can be found. Assuming that the blobs are circular, we propose to reproduce their diameter distribution by the mathematical expression given by Eq. (6). Taking the log of this expression twice, it becomes:

$$\ln\left(\left|\ln\left(\frac{e_2(d),_d}{e_2(0),_d}\right)\right|\right) = q\ln(d) - q\ln(D_c) - \ln(q) \quad (14)$$

Eq. (14) indicates a linear evolution in a log x log frame of the logarithm of the distribution $e_2(d),_d$ with scale $d$. By replacing $e_2(0),_d$ by $e_2(d_{pc}),_d$, the relevance of the model expressed by Eq. (14) is evaluated on measurements. An example is shown in Fig. 10 for the result obtained at $z = 12.1$ mm.

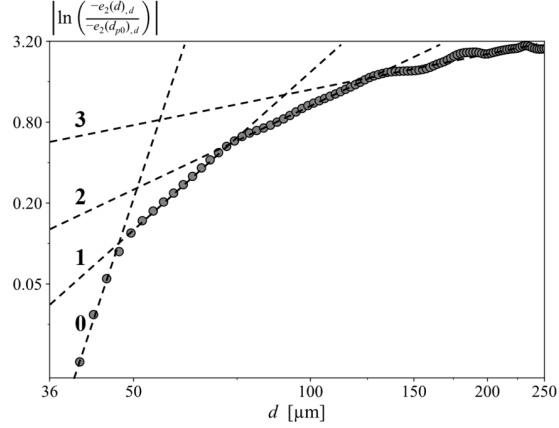

**Figure 12:** Illustration of Eq. (14) at $z = 12.1$ mm: $\ln(\ln(-e_2(d),_d/-e_2(d_{p0}),_d) = f(\ln(d))$
Dash lines indicate linear behaviors of the curve

Fig. 12 shows four regions where the curve is linear. The model proposed by Eq. (14) is therefore relevant and reports the existence of several families of blobs, one per linear region. Region 0 (see Fig. 12) is supported by only a few points and is not observed at all $z$ positions. Furthermore, examination of the images reveals no blobs of the size of the scales covered by this region. Region 1 corresponds to blobs beading at the ligament interface such as those shown in white in Fig. 9. Region 2 covers the scales of blobs formed on the body of ligaments such as those shown in light grey in Fig. 9. Region 3 covers the scales of blobs formed at the base of ligaments such as those shown in dark grey in Fig. 9.

Several families of blobs are therefore identified, each of which may be representative of drops in the production phase. This possibility needs to be considered by proposing a model that can identify several families of blobs at the same time. As a first approach, we propose a model to identify two blobs populations of two consecutive regions by introducing two components in the derivative of the scale distribution, namely:

$$-e_2(d),_d = N_{ci}f_{0ci}(d) + N_{c(i+1)}f_{0c(i+1)}(d) \quad (15)$$

where $i = 1$ or 2 refers to the analysis of regions 1 and 2 or 2 and 3, respectively. In Eq. (15), $f_{0ci}(d)$ is given by (see Eq. 6):

$$f_{0ci}(d) = \frac{1}{D_{ci}} \frac{q_i^{(1-1/q_i)}}{\Gamma\left(\frac{1}{q_i}\right)} exp\left(-\frac{1}{q_i}\left(\frac{d}{D_{ci}}\right)^{q_i}\right) \quad (16)$$

The model expressed by Eq. (15) is applied as follows.

Six parameters insuring the best fit of $-e_2(d)_{,d}$ with Eq. (15) need to be determined: $N_{ci}$, $q_i$ and $D_{ci}$ for the first component, and, $N_{c(i+1)}$, $q_{(i+1)}$ and $D_{c(i+1)}$ for the second component. By introducing in the model a value for $-e_2(0)_{,d}$ (see Eq. (6)), one of the parameter can be expressed as a function of the five others, and the number of parameters to be determined reduces to five. The model is applied for a given scale range, i.e., $[d_1; d_2]$, covering the couple of regions of interest and determined on a curve like the one shown in Fig. 12. $-e_2(0)_{,d}$ is taken equal to the experimental value of $-e_2(d_1-\Delta d)_{,d}$. The parameter search routine is the Python *scipy.optimize.minimize* routine. Once the parameters insuring the best fit with the measurements are determined, the blob diameter distribution of each component $f_{0si}(D)$ and $f_{0s(i+1)}(D)$ can be determined (Eq. (8)), as well as the diameter $D_{si}$ and $D_{s(i+1)}$ (Eq. (9)) and the blob number $N_{si}$ and $N_{s(i+1)}$ (Eq. (12)). The numbers $N_{si}$ and $N_{s(i+1)}$ do not actually represent the real number of blobs since their values depend on the parameter $L_c$ that was arbitrarily taken equal to 1 μm in the present analysis. $N_{si}$ and $N_{s(i+1)}$ are proportional to the real blob number in the same way for all cases allowing performing comparisons.

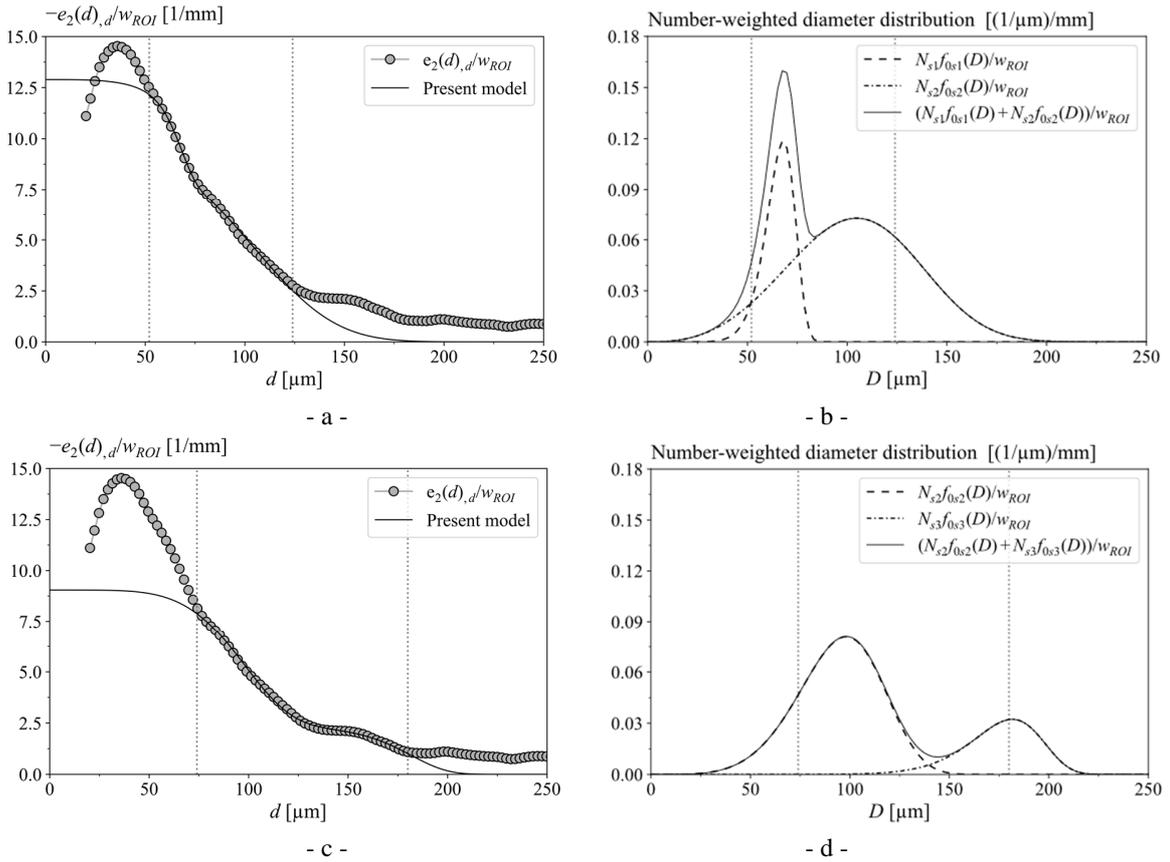

**Figure 13:** Fit of $-e_2(d)_{,d}$ (left graphs) and corresponding weighted blob distributions (right graphs, Eq. (11), each component and sum) $z = 12.1$ mm
- a and b: regions 1 and 2, scale range [52 μm; 124 μm]
- c and d: regions 2 and 3, scale range [74 μm; 180 μm]

The results of the applications of the model for regions 1 and 2 ($i = 1$) at $z = 12.1$ mm are shown in Fig. 13-a and 13-b. (The distributions are plotted per unit of space.) In each plot, the vertical dash lines delimit the analyzed scale range [$d_1$; $d_2$]. Figs. 13-a reports that the model proposes a very acceptable fit of $-e_2(d)_{,d}$. The number-weighted diameter distributions in Fig. 13-b show that the blob populations from regions 1 and 2 are very distinct with a peak diameter equal to 70 µm and 100 µm for population 1 and 2, respectively. Furthermore, the width of the distribution of region 1 blobs is far less than the width of the distribution of the region 2 blobs. The surface area delimited by each number-weighted blob diameter distribution is equal to the number of blobs of each component and the sum $N_{s1} + N_{s2}$ represents the total blob number. In the case shown in Fig. 13-b, region 2 contains 73% of the total blob number.

Figures 13-c and 13-d show the application of the model for regions 2 and 3 ($i = 2$) at $z = 12.1$ mm. Again, we note that the model returns a good fit of the distribution in the selected scale interval [$d_1$; $d_2$]. As for the previous application, regions 2 and 3 show distinctive blob populations and the global distribution is bi-modal. However, the two distributions overlap less than for the first case. The peak diameter of the population is equal to 100 µm and 190 µm for regions 2 and 3, respectively. The distribution width for region 2 blob population is greater than for region 3 blob population. Region 2 contains 74% of the total blob number.

Although quite similar, the two blob diameter distributions in region 2 reported by the two analyzes show a difference in width: the one in Fig. 13-b is larger than the one in Fig. 13-d and extends to the peak diameter of the region 3 blob diameter distribution, i.e., 190 µm. This difference is due to the fact that the model extends the distribution beyond $d_2$ insofar as the derivative $-e_2(d_2)_{,d}$ is never zero. Thus, the diameter distribution of blobs in region 2 obtained by analyzing regions 1 and 2 includes some small region 3 blobs. Note that among the blobs of region 3, it is not excluded that some small ones produce drops as Fig. 9 suggests. Thus, we believe that the analysis of regions 1 and 2 reports an estimation of the diameter distribution of the blobs that will become drops.

The analysis of regions 1 and 2 was performed at each $z$ positions. The number-weighted diameter distributions of the blobs of each region as well as their sum are plotted in Fig. 14. On each graph, the scale interval [$d_1$; $d_2$] is indicated by vertical dash lines. The results are divided by the ROI width to allow comparison of the graphs from one position to another. We must add that at the 23.5 mm and 25.6 mm $z$ positions the region 1 blobs was never found. Therefore, for these positions, the region 2 was analyzed only by using a model reduced to a single component. The scale intervals [$d_1$; $d_2$] for these two mono-modal applications were defined on the basis of the evolution of $d_1(z)$ and $d_2(z)$ built from the preceding positions. (Linear spatial evolutions were found for these two scales.)

Figure 14 shows that the diameter distribution of the blobs of region 1 varies little with the position. Indeed, the peak position of the distribution remains rather the same. On the other hand, the number of elements in this population increases significantly from $z = 3.36$ mm to 12.1 mm and then collapses between $z = 12.1$ mm and 21.3 mm. As mentioned above, this population is not observed anymore at the two farthest $z$ positions. On the other hand, the diameter distribution of the blobs of region 2 varies continuously with the position: the peak diameter and width both increase. Furthermore, the distribution first swells from $z = 3.36$ mm to 21.3 mm and then deflate from $z = 21.3$ mm to 25.6 mm.

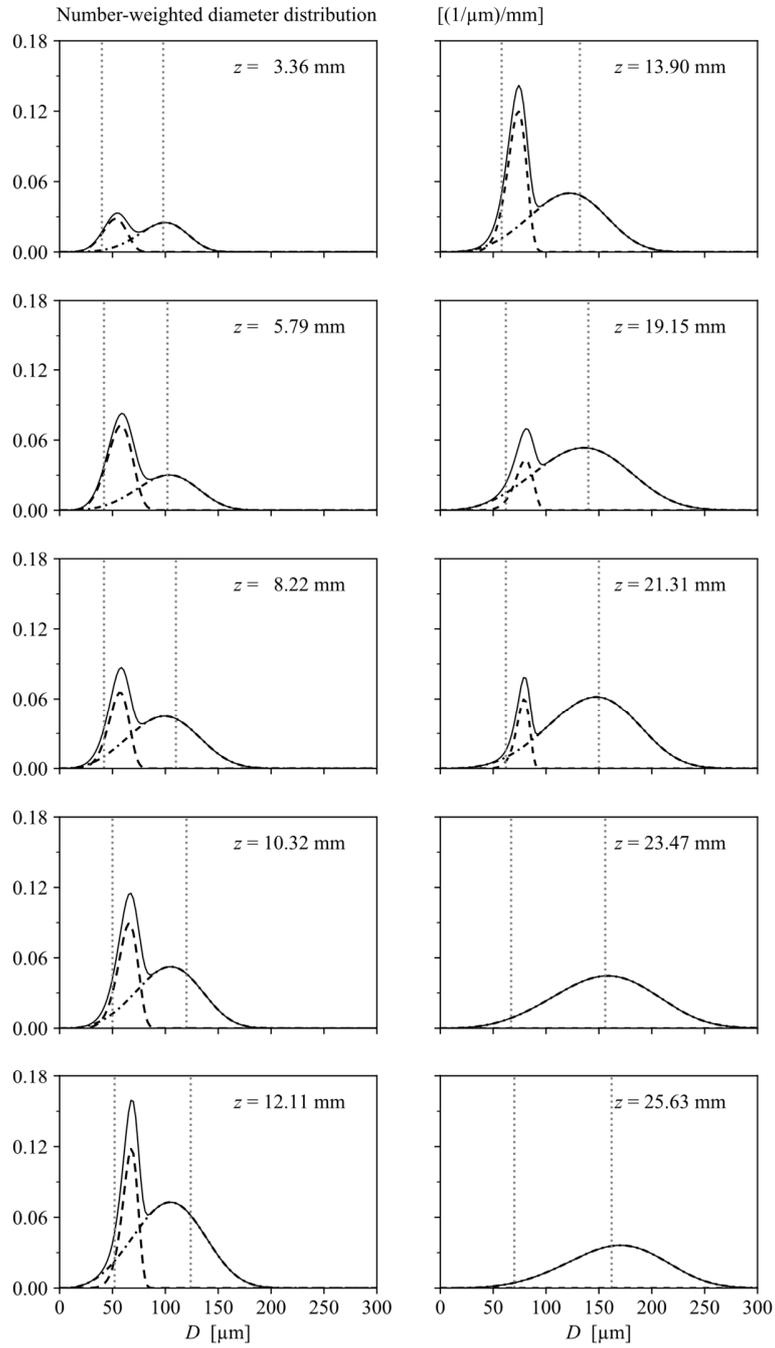

- - - $N_{s1}f_{0s1}(D)/w_{ROI}$    -·-·- $N_{s2}f_{0s2}(D)/w_{ROI}$    ——— $(N_{s1}f_{0s1}(D)+N_{s2}f_{0s2}(D))/w_{ROI}$

**Figure 14:** Components $N_{si}f_{0si}(D)/w_{ROI}$ and sum for each position
(vertical dash lines delimit the analyzed scale interval)

The study of the parameters of the blob diameter distributions as a function of the $z$ position assesses the possibility of proposing a model for these distributions. The evolution of these parameters with $z$ are plotted in Fig. 15. The following points are noted.

The parameters $q_1$ and $q_2$ report linear spatial evolutions. The increase of $q_1$ denotes a distribution component with an increasingly smaller dispersion, the dispersion being a non-dimension distribution width. On the other hand, the rather constant $q_2$ denotes a distribution component with a rather constant dispersion. Note that both $q_1$ and $q_2$ are above 2.5 which indicates that each component can be considered as symmetric.

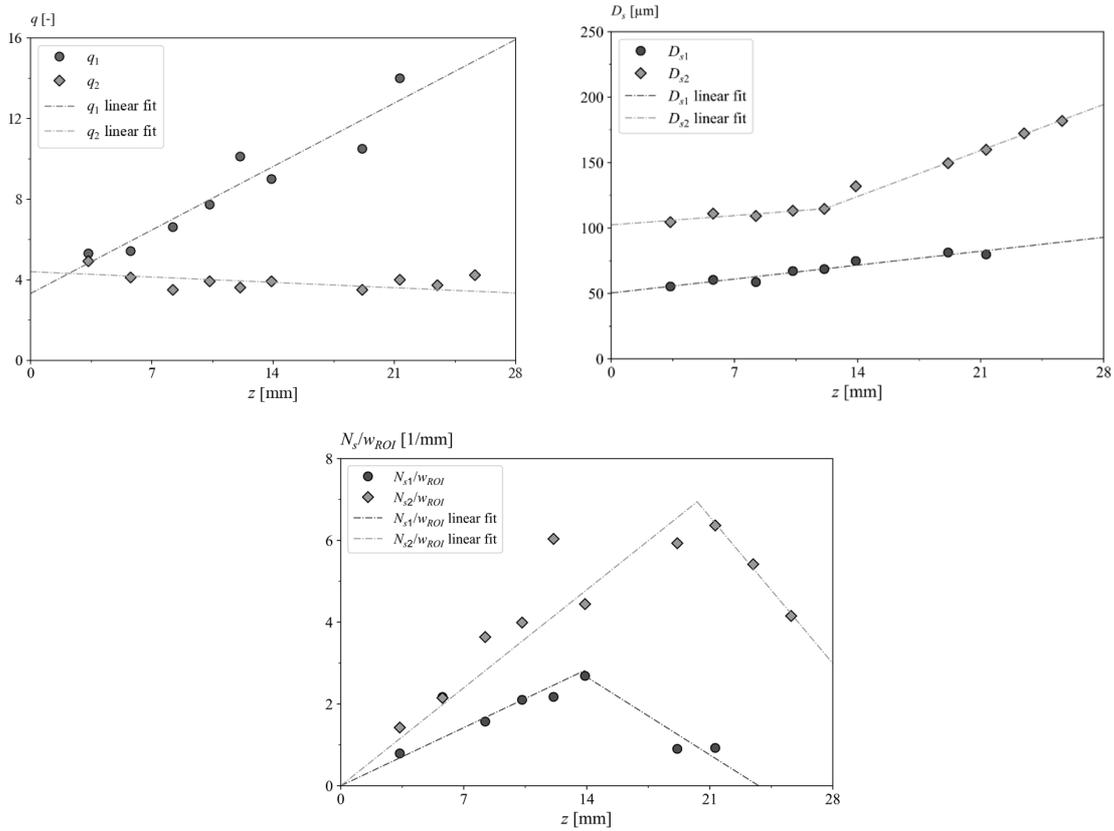

**Figure 15:** The parameters of the model.

$D_{s1}$ and $D_{s2}$ also report a linear increase but, with a change of slope for $D_{s2}$. The slopes of the mean diameter $D_{s2}$ linear increase are slightly greater than the one of $D_{s1}$. Note here that the $D_{s2}$ obtained at $z = 23$ mm and 25.6 mm align perfectly well with those obtained at smaller $z$.

The spatial evolutions of the number $N_{s1}/w_{ROI}$ and $N_{s2}/w_{ROI}$ are not monotonous: they both increase and then decrease with the position $z$. Note that the position at which $N_{s1}/w_{ROI}$ decreases ($z = 14$ mm) is the same at which the slope of diameter $D_{s2}$ changes. The total number of blobs, i.e. $N_{s1}/w_{ROI} + N_{s2}/w_{ROI}$, is highest in the region [14 mm; 21 mm]. This information indicates that the volume of liquid contained in the ligament structures that make up the textural atomization process is at its maximum in this position interval. The spatial evolutions of the model parameters are clear and therefore suggest the possibility of building models to predict the spatial evolution of the diameter distribution of the blobs composing the ligaments of textural atomization processes.

If we accept the idea that the diameter distributions of the blobs are an indication of the diameter distributions of the drops produced by these blobs, the results of the analysis indicate that the average size of these drops increases with distance from the injector and that their local production rate is maximum in a region located here between 14 mm and 21 mm from the injector. It is interesting to note that the increase in diameter of the textural drop population is identical to that of drops resulting from the primary atomization process described by Wu and Faeth (1993). This gives credit to the idea that the ligament structures analyzed here seem to be those resulting directly from the impact of turbulence carried by the liquid flow.

**5. Conclusion**

The method presented in this article for the description of textural atomization liquid ligaments is convincing and brings back little-known information on this type of process. An important part of this success is linked to the association of high-quality experimental images with improved image processing and image analysis measurement techniques. Evaluated on a synthetic object, the improved image analysis procedure increases the measurement accuracy in the small-scale range which is important when we are interested in describing small-scale mechanisms such as textural atomization processes.

The measurement in question is that of the scale distribution of the ligament structures that make up the textural atomization process. It provides information on the length, size, deformation and number of ligaments involved in the process. So, for example, it can be used to identify the area where the process is most intense in terms of volume of involvement, and where it disappears. Moreover, combined with a mathematical description built on simple objects (cylinders and spheres), the measurement leads to an estimate of the size distribution of the blobs that structure the deformation of the ligaments and that can be seen as drops in formation.

Within the scope of this article, the method is applied to one experimental case corresponding to a subcritical injection operation condition similar to those encountered in rocket engine coaxial gas-assisted atomization. The analysis is performed as a function of the distance from the injector.

It is found that the blob size distribution is bi-modal revealing the presence of two atomization processes in the textural mechanism. This bi-modal character is due the fact that the textural ligaments show textural and structural deformations: Textural deformations are associated with small-sized blobs, whereas structural deformations are associated with larger blobs. The production of the textural ligaments and the development of textural deformations of these ligaments are the result of two mechanisms impacting the spray produced by the textural atomization process in the present application.

The analysis provides further information on these mechanisms. For example, blob production by ligament texture ceases at a certain distance, at which point the blob diameter distribution becomes monomodal, with only the blobs carried by the ligament structure remaining. Furthermore, the position at which the textural atomization stops is also determined by the analysis.

We see here that the proposed description method and the analysis extracted from it provide new, unusual information on atomization processes which open up new areas of study of these mechanisms.

Finally, it should be noted that the mathematical tool used in the development of the method associates a set of parameters with the identified blob distributions. The variations of these parameters as a function of distance from the injector are sufficiently organized to allow us to build a model for the blob diameter distribution. Assuming that the analyzed blobs correspond to drops in formation, this model could be used for the drop diameter distribution produced by textural atomization. Of course, this family of drops does not include the small elements that can be formed when blobs detach. This type of model can be very useful, for example, to enable numerical simulations of atomization to replace fine textural ligaments with a family of spherical drops whose diameter distribution is known.